\documentclass{elsart5p}

\usepackage{graphics}
\usepackage{graphicx}
\usepackage{amssymb}

\newcommand{\tise}{1\textit{T}-TiSe$_2$}

\begin{document}

\begin{frontmatter}

\title{Temperature dependence of the excitonic insulator phase model in 1\textit{T}-TiSe$_2$}

\author[A]{C. Monney},
\ead{claude.monney@unine.ch}
\author[B]{H. Cercellier},
\author[A]{C. Battaglia},
\author[A]{E.F. Schwier},
\author[A]{C. Didiot},
\author[A]{M. G. Garnier},
\author[A]{H. Beck},
\author[A]{P. Aebi}

\address[A]{%
Institut de Physique, Universit\'e de Neuch\^atel, CH-2000 Neuch\^atel, Switzerland}%
\address[B]{%
Institut N\'eel, CNRS, F-38042 Grenoble, France}%



\begin{abstract}

Recently, detailed calculations of the excitonic insulator phase model adapted to the case of 1\textit{T}-TiSe$_2$ have been presented.
Through the spectral function theoretical photoemission intensity maps can be generated which are in very good agreement with experiment  [Phys. Rev. Lett. {\bf 99}, (2007) 146403]. In this model, excitons condensate in a BCS-like manner and give rise to a charge density wave, characterized by an order parameter. Here, we assume an analytical form of the order parameter, allowing to perform temperature dependent calculations. The influence of this order parameter on the electronic spectral function, to be observed in photoemission spectra, is discussed. The resulting chemical potential shift and an estimation of the resistivity are also shown.

\end{abstract}


\end{frontmatter}

\section{Introduction}

The transition metal dichalcogenide \tise\ is a layered compound exhibiting a commensurate (2x2x2) charge density wave (CDW) at low temperature\cite{DiSalvo}. In transport measurements, its signature is a strong anomaly in the resistivity, peaking slightly below the transition temperature $T_c=200$K. The CDW is accompanied by a periodic lattice distortion involving small atomic displacements. At the same time, a zone boundary phonon softens, being a central ingredient to conventional theories of structural transitions\cite{HoltPhonon}. High resolution angle-resolved photoemission spectroscopy (ARPES) measurements brought a deeper insight into the electronic properties of \tise\ by clearly revealing the appearance of new backfolded bands, characterisitc of a new periodicity in the system\cite{Pillo,Rossnagel,Kidd}.

The origin of the CDW can hardly be explained by the usual nesting mechanism\cite{nesting}, because there are notably no large parallel Fermi surface portions\cite{Johannes}.
Currently, the best candidates are a band Jahn-Teller effect\cite{HughesBJT} and the excitonic insulator phase. Recently, the latter scenario has been strongly supported by comparison of ARPES data with theoretical photoemission intensity maps \cite{CercellierPRL}. These calculations are based on the excitonic insulator phase model \cite{Kohn,Jerome}, which has been adapted to the particular case of \tise \cite{MonneyPRB}.
The excitonic insulator phase may occur in a semimetallic or semiconducting system exhibiting a small (negative respectively positive) gap. Indeed, for a low carrier density, the Coulomb interaction is weakly screened, allowing therefore bound states of holes and electrons, called excitons, to build up in the system. If the binding energy $E_B$ of such pairs is larger than the gap $E_G$, the energy to create an exciton becomes negative, so that the ground state of the normal phase becomes unstable with respect to the spontaneous formation of excitons. At low temperature, these excitons may condense into a macroscopic coherent state in a manner similar to Cooper pairs in conventional BCS superconductors. Exciton condensation may lead to the formation of CDW of purely electronic origin (not initiated by a lattice distortion), characterized by an order parameter $\Delta$. 
To our knowledge, \tise\ is the only presently known candidate for a low temperature phase transition to the excitonic insulator state without the influence of any external parameters other than temperature. Indeed, as pressure is increased above 6 kbar on TmSe$_{0.45}$Te$_{0.55}$ ( controlling the gap size and thus the energy necessary to create excitons), a transition to an insulating phase happens, whose origin can also be explained with exciton condensation \cite{Wachter}. In this context Bronold and Fehske proposed an effective model for calculating the phase boundary of a pressure-induced excitonic insulator, in the spirit of a crossover from a Bose-Einstein to a BCS condensate \cite{Bronold}.

Here, we study the temperature dependence of the excitonic insulator phase. This is achieved by choosing a simple analytical form for its order parameter and inserting it into the results of our previous calculations \cite{MonneyPRB}. 

\section{Results and Discussion}

\begin{figure}[ht]
\begin{center}
\includegraphics[width=8.7cm]{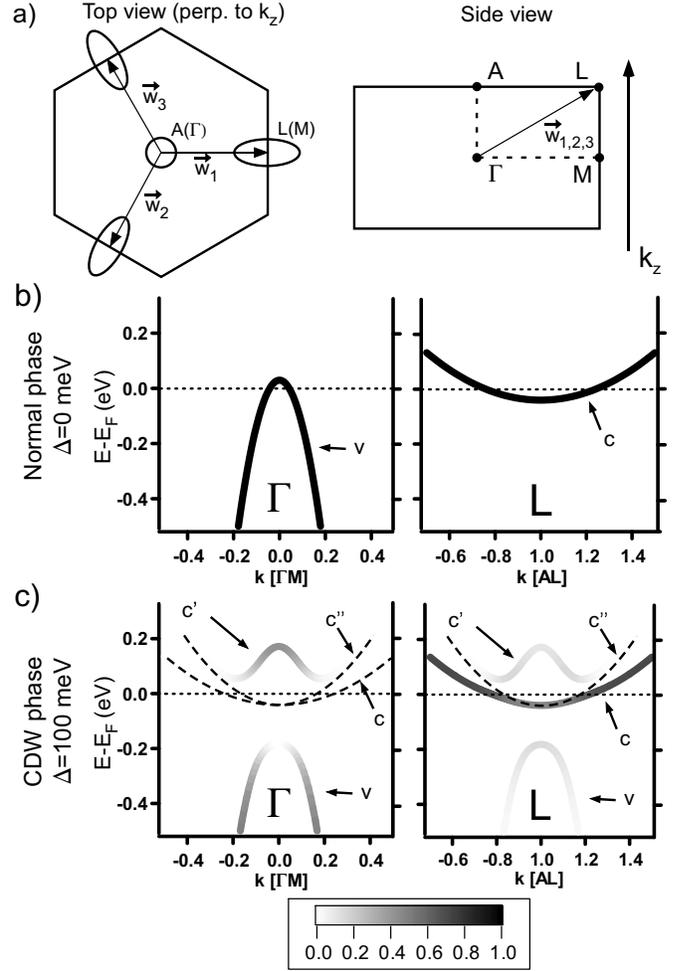}
\end{center}
\caption{(a) On the left, schematic picture of the Fermi surface (in the plane perpendicular to $k_z$) considered in the model, placed on the Brillouin zone of \tise. On the right, side view of the Brillouin zone showing high symmetry points. Graphs (b) and (c) show band dispersions with their spectral weight (photoemission spectra) calculated with the spectral function at $\Gamma$ and $L$ along the high symmetry direction $\Gamma M$ and $AL$ (long axis of the electron pockets) for an order parameter $\Delta=0$ meV and $\Delta=100$ meV respectively. The spectral weight carried by the band is indicated in grayscale. In Graphs (c), the thin dashed lines indicate the position of bands having negligibly small spectral weight.} \label{Fig1}
\end{figure}

The electronic structure of \tise\ near the Fermi energy $E_F$ is composed of three Se$4p$-derived valence bands at the $\Gamma$ point (center of Brillouin zone) and three Ti$3d$-derived conduction bands distributed among the three symmetry equivalent $L$ points (zone boundary). There is a slight overlap of $\sim 70$ meV, such that \tise\ has a semimetallic character. This issue is still controversial among the ARPES community, but a recent infrared study confirmed the semimetallicity \cite{Li}. In an ionic picture, Ti [Ar]$3d^24s^2$ gives all its valence electrons to the two neighbouring Se [Ar]$3d^{10}4s^24p^4$, leaving a system with an empty $d^0$ shell. In our model, we consider only the topmost valence band (the other two do not cross the Fermi energy and play a minor role) and the three conduction bands. The valence band and the three symmetry equivalent conduction bands give rise to a hole pocket at $\Gamma$ and electron pockets at $L$ (see Fig. \ref{Fig1}(a)). Their band dispersions, $\epsilon_v$ for the valence band and $\epsilon_c^i$ ($i=1,2,3$) for the conduction bands, have been approximated by a parabolic form which describes them well near their extrema, in agreement with ARPES experiment \cite{CercellierPRL}.

Our Hamiltonian is composed of a one-electron part $H_0$, containing the one-electron energies, and a Coulomb interaction part $W$, which represents the electron-hole interaction between the valence and the conduction bands. Below $T_c$, this interaction allows electron-hole pairs, i.e. excitons, to form a condensate described by the order parameter $\Delta$. The calculated Green's functions $G_v$ and $G_c$ describing the bandstructure at $\Gamma$ and $L$ respectively have the following forms \cite{MonneyPRB}
\begin{eqnarray} 
G_v(\vec{p},z)&=&
\frac{1}{\mathcal{D}(\vec{p},z)}
\cdot\prod_i(z-\epsilon_c^i(\vec{p}+\vec{w}_i)),\nonumber\label{eqn_valGreenDenom}\\
G_c^i(\vec{p},z)&=&\frac{1}{\mathcal{D}(\vec{p},z)}\cdot\Big((z-\epsilon_v(\vec{p}))\prod_{j\neq i}(z-\epsilon_c^j(\vec{p}+\vec{w}_j))\nonumber\\&&-|\Delta|^2\sum_{m,j\neq i}|\varepsilon_{ijm}|(z-\epsilon_c^m(\vec{p}+\vec{w}_j))\Big)\nonumber\label{eqn_condGreenDenom}
\end{eqnarray}
($\varepsilon_{ijm}$ is the permutation symbol). The vectors $\vec{w}_i$, called spanning vectors, correspond to the distance between $\Gamma$ and the $L$ points. The denominator $\mathcal{D}$, common to both Green's functions, is
\begin{eqnarray}\label{eqn_denominator}
\mathcal{D}(\vec{p},z)&=&(z-\epsilon_v(\vec{p}))\prod_i(z-\epsilon_c^i(\vec{p}+\vec{w}_i))\nonumber\\&-&|\Delta|^2\sum_i\prod_{j\neq i}(z-\epsilon_c^j(\vec{p}+\vec{w}_j)).\nonumber
\end{eqnarray}
The zeroes of this denominator give the renormalized band dispersions, which depends on the order parameter $\Delta$. They are common to $\Gamma$ and $L$, i.e., for  $G_v$ and $G_c$, as one expects for a CDW characterized in our case by the spanning vectors $\vec{w}_i$. The spectral function, $A(\vec{p},\Omega)=-$Im$[ G(\vec{p},\Omega+i\delta)]/\pi$, describes the one-electron spectrum, essential for our purposes. It provides us with the spectral weight (SW) carried by the dispersions in the process of photoemission. Fig. \ref{Fig1}(b) and (c) present calculated photoemission spectra at $\Gamma$ and $L$ along the high symmetry direction $\Gamma M$ and $AL$ (long axis of the electron pockets) for an order parameter $\Delta=0$ meV (normal phase) and $\Delta=100$ meV (CDW phase) respectively. The SW of the dispersions is indicated in grayscale. On Fig. \ref{Fig1} (c), the bands indicated by the thin dashed lines have a negligibly small SW, so that they do not appear on these grayscale graphs.
We immediately see that the similarity of the bands at $\Gamma$ and $L$ is only approximate. At $\Gamma$, as the order parameter increases, the lower part of the valence band $v$ shifts to higher binding energies, while its top $c'$ shifts above $E_F$, opening thereby a gap. Some SW is then tranferred from $v$ to $c'$. At $L$, with an increasing order parameter, the conduction band $c$ does not move but looses SW in favor of the backfolded valence band $v$ and the new band $c'$ (which are the same as those appearing at $\Gamma$).
\\

\begin{figure}[ht]
\begin{center}
\includegraphics[width=6cm]{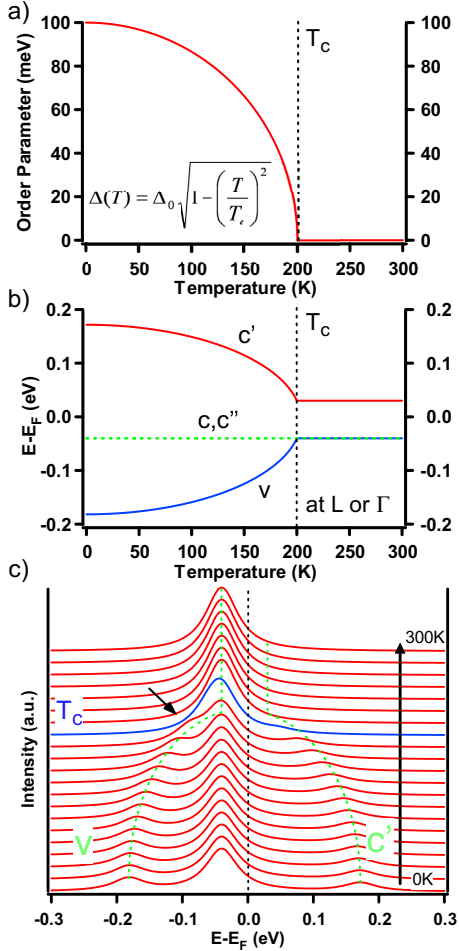}
\end{center}
\caption{(a) BCS-like temperature dependence chosen for the order parameter. (b) Position of the bands at $\Gamma$ and $L$ as a function of a temperature (for the above order parameter). (c) Evolution of the broadened spectra (see text) at $L$ under the effect of temperature.} \label{Fig2}
\end{figure}

In order to introduce temperature effects in the model in a simple way, we now choose a particular form for the order parameter $\Delta(T)=\Delta_0\cdot\sqrt{1-\left(T/T_c\right)^2}$ where $\Delta_0$ is its value at $T=0$K. This function is drawn in Fig. \ref{Fig2}(a) for $\Delta_0=100$ meV and $T_c=200$K. It is similar to a BCS order parameter, displaying a steep decrease at $T_c$ and a saturation for $T\rightarrow 0$K. Introducing this order parameter into the denominator of $\mathcal{D}$ and calculating its zeros provides us with the curves of Fig. \ref{Fig2} (b). They show the temperature dependence of the band positions at $\Gamma$ and $L$. Below the transition temperature, the backfolded valence band $v$ and conduction band $c'$ shift away from their normal phase positions, exhibiting a behaviour very similar to the shape of the order parameter. At the same time, the conduction band $c$ and a symmetry equivalent version $c''$ stay at their inital position. 
Considering also SWs and broadening the $\delta$-like peaks with a finite width of 30 meV (for presentation purposes), Fig. \ref{Fig2} (c) displays over a wide temperature range spectra at $L$ (not equivalent to $\Gamma$ due to the SWs), where the excitonic effects are the most spectacular. Here, below $E_F$, one sees that the evolution of the backfolded valence band $v$ is characteristic of the temperature dependence of the order parameter (as the backfolded conduction band $c'$, which is however not accessible to photoemission, since the states are unoccupied). These calculated spectra highlight how the real (experimental) temperature behaviour of the order parameter can be extracted from ARPES data. One sees that the situation is particularly delicate when the order parameter is small, since the peak of the backfolded valence band $v$ merges with the peak of the conduction band (see arrow on Fig. \ref{Fig2} (c)).

\begin{figure}[ht]
\begin{center}
\includegraphics[width=9cm]{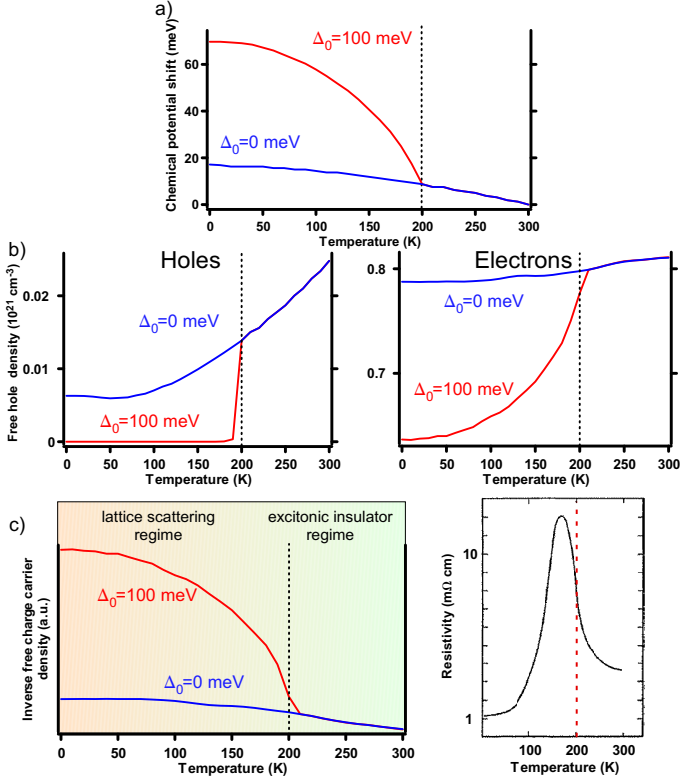}
\end{center}
\caption{Compared behaviours for the cases of a simple semi-metallic system (i.e. $\Delta_0=0$ meV, in blue) and an excitonic insulator (i.e. $\Delta_0=100$ meV, in red). (a) Temperature dependence of the chemical potential. (b) Temperature depence of the free hole (left) and electron (right) density. (c) Left graph: inverse free carrier density approximating the resistivity (see text for explanation concerning the two regimes). Right graph: the measured inplane resistivity of \tise\cite{Pillo}} \label{Fig3}
\end{figure}

From the condition of conservation of occupied electronic states (weighted with the SW), one can also compute the temperature dependence of the chemical potential $\mu$. 
Technically, we only take into account states down to $-1.0$ eV, due to the parabolic approximation of the bands around their extrema. Then, we compute the number of occupied electronic states $n_\mathrm{occ}$ at $T=300$K (taking into account the SW of the dispersions) for getting a reference value. Then, decreasing the temperature gradually (and possibly increasing the order parameter), we adapt the chemical potential so that the number of occupied electronic states remains constant $n_\mathrm{occ}(T<300\mathrm{K})=n_\mathrm{occ}(T=300\mathrm{K})$.
This has been done for two differents cases. First we fix $\Delta_0=0$ meV, which means that we look at a normal system, exhibiting no transition. Then we fix $\Delta_0=100$ meV, as before, for the excitonic insulator system. Fig. \ref{Fig3} (a) shows the behavior of the chemical potential. One sees that without excitonic effects, a system having the modelized configuration of \tise\ undergoes already a chemical potential shift of about 12 meV over the range of 300K due to the change in the thermal occupation of electronic states. If an excitonic phase transition with $\Delta_0=100$ meV sets in, a drastic change happens around $T_c$ and the chemical potential shift increases up to nearly 60 meV. Such a large effect should be visible in ARPES, but it must be emphasized that this holds only for the simplified \tise electronic bandstructure of this model (in particular without considering the other Se$4p$-derived valence bands).

Knowing the band dispersions around $E_F$ and their shift due to the combined effect of the chemical potential and the order parameter, it is also possible to estimate the free carrier density $n=n_{\mathrm{hole}}+n_{\mathrm{electron}}$ in the system, composed of holes in the valence band and electrons in the conduction bands. 
In Fig. \ref{Fig3} (b) (left), in comparison with the normal metal ($\Delta_0=0$ meV), the excitonic insulator ($\Delta_0=100$ meV) displays a strong free hole density decrease below $~T_c$. This can be explained with Fig. \ref{Fig1} (b) and (c). As the order parameter increases from a zero value, a gap opens at $\Gamma$ and $\mu$ remains in the conduction band so that the hole contribution to conduction vanishes. 
In Fig. \ref{Fig3} (b) (right), the free electron density of the excitonic insulator displays also a strong decrease below $T_c$. Indeed, at $L$, the bottom of the conduction band $c$ looses SW, reducing the free electron carrier density. In parallel, the chemical potential moves upward in the conduction band (this shift is nonetheless smaller than the gap produced by the order parameter), making available new states in the conduction band $c$ with higher SW than those at its bottom. However, this is not sufficient to counterbalance the previous effect.

In transport measurements, the CDW has a strong signature, raising the resistivity $\rho$ below $T_c$ as in an insulator. Having now an approximate behaviour of the free charge carrier density as a function of temperature, we can also estimate the resistivity in the Drude theory by $\rho=m/ne^2\tau$, with $\tau$ the relaxation time and $m$ the effective mass. Fig. \ref{Fig3} (c) shows $1/n$, one ingredient of $\rho$, again for $\Delta_0=0$ meV and $\Delta_0=100$ meV. In comparison to the normal system ($\Delta_0=0$ meV) which displays a nearly constant $1/n$, the excitonic insulator ($\Delta_0=100$ meV) exhibits a strong increase of $1/n$ below $T_c$, as expected. Compared to the measured inplane resistivity shown in Fig. \ref{Fig3} (c) (right), one sees a relatively good qualitative agreement above 170K (indicated as the excitonic insulator regime in Fig. \ref{Fig3} (c)). At low temperature, our crude estimation of the resistivity, $\rho\propto 1/n$, does not take into account the scattering by the lattice represented by the relaxation time $\tau$ in the Drude theory. Indeed, for low temperatures, $\tau$ increases as a power of $T$, compensating thereby the decreasing $n$. This generally happens well below the Debye temperature, evaluated as $\Theta_D\cong200$K \cite{Baranov} (indicated as the lattice scattering regime in Fig. \ref{Fig3} (c)).
\\

Of course, the order parameter of the excitonic insulator model could be computed directly in the framework of the model, since it obeys to a gap equation similar to that of BCS theory. The chemical potential would be obtained as well in a self-consistent way. Notwithstanding, such a computation is demanding and goes beyond the discussion of the present study. 

\section{Conclusions}

The temperature dependence of the excitonic insulator phase model adapted to 1\textit{T}-TiSe$_2$  \cite{CercellierPRL, MonneyPRB} has been studied. This was done in a simple way by assuming a given temperature dependent order parameter. Then, photoemission spectra were calculated as a function of temperature, with the help of the spectral function derived from our model. They inform us about the signature of the order parameter in photoemission and indicate how its real (experimental) behaviour can be extracted from ARPES experiments. Moreover the upturn in the measured resistivity of \tise\ was explained in terms of the excitonic transition.

\section{Acknowledgments}

This work was supported by the Fonds National Suisse pour la Recherche Scientifique through Div. II and MaNEP.

\end{document}